\documentclass[amsmath,amssymb]{revtex4}
\usepackage{graphicx}
\usepackage{dcolumn}
\usepackage{bm}

\newcommand{\gsim}{\raisebox{0.2ex}{$\ > \kern -1.05em%
        \raisebox{-1.1ex}{$\sim$}\ $}}  
\begin{document}

\title{Effective potential analytic continuation approach \\
for real time quantum correlation functions 
involving nonlinear operators
}

\author{Atsushi Horikoshi$^{1,2}$}
\altaffiliation{Present address: 
Department of Chemistry, Graduate School of Science,
Kyoto University, Kyoto 606-8502, Japan}
 \email{horikosi@kuchem.kyoto-u.ac.jp}
\author{Kenichi Kinugawa$^{2}$}%
 \email{kinugawa@cc.nara-wu.ac.jp}
\affiliation{
$^{1}$Japan Science and Technology Agency 
\\and\\
$^{2}$Department of Chemistry, Faculty of Science, 
Nara Women's University,
 \\
Nara 630-8506, Japan}

\date{\today}

\begin{abstract}
We apply the effective potential analytic continuation (EPAC) method 
to the calculation of real time quantum correlation functions 
involving operators nonlinear in the position operator $\hat{q}$.
For a harmonic system the EPAC method provides the exact correlation function
at all temperature ranges, 
while the other quantum dynamics methods,
the centroid molecular dynamics and the ring polymer molecular dynamics,
become worse at lower temperature. 
For an asymmetric anharmonic system,
the EPAC correlation function is  
in very good agreement with the exact one at $t=0$.
When the time increases from zero, 
the EPAC method gives 
good coincidence with the exact result at lower temperature.
Finally, we propose a simplified version of the EPAC method
to reduce the computational cost required for the calculation of 
the standard effective potential.
\end{abstract}
\maketitle
\section{INTRODUCTION}
\hspace*{\parindent}
The imaginary time path integral \cite{fh}
has provided a useful framework 
suitable for numerical analyses of
quantum statistical-mechanical systems.
Most of the static properties of quantum systems
can be calculated by means of the 
path integral Monte Carlo (PIMC) or 
path integral molecular dynamics (PIMD) technique \cite{bt,ce}.
However, 
it is not straightforward to apply the PIMC/PIMD methods   
to computing dynamical properties such as 
the real time quantum correlation function 
$\langle \hat{A}(t)\hat{B}(0)\rangle_{\beta}$.
This is because 
it is nontrivial to construct real time quantities  
from a finite number of imaginary time data obtained numerically \cite{tb}.
To overcome such difficulty, 
a number of promising methods of 
numerical analytic continuation based on 
the maximum entropy method have been proposed
and applied to various many-body systems \cite{si,gb,rr,rr1}.

\par
Recently,
the centroid molecular dynamics (CMD) method \cite{cv},
the ring polymer molecular dynamics (RPMD) method \cite{cm},
and 
the effective potential analytic continuation (EPAC) method \cite{hk1}
have been proposed
as new quantum dynamics methods to calculate 
real time quantum correlation functions 
at finite temperature.
Both the CMD and the RPMD are the methods to calculate 
the canonical (Kubo-transformed) correlation function \cite{kubo}
$\langle \hat{A}(t)\hat{B}(0)\rangle_{\beta}^{\rm can}$
by means of molecular dynamics techniques.
On the other hand, the EPAC is a method 
to obtain the real time quantum correlation function 
$\langle \hat{A}(t)\hat{B}(0)\rangle_{\beta}$ 
by means of the effective action formalism \cite{ep,riv,ps} 
and an analytic continuation procedure \cite{bell}.  
It has been shown analytically that
all these methods are exact in harmonic systems
for the real time quantum correlation functions 
of a linear function of the position operator,
$\hat{A}=\hat{B}=\hat{q}$ \cite{jv,cm,hk2}.
However, for nonlinear operators 
such as $\hat{A}=\hat{B}=\hat{q}^{n}$,
it is nontrivial whether these quantum dynamics methods 
yield the exact result
even in harmonic systems \cite{jv,cm}.
This is the {\it nonlinear operator problem} in quantum dynamics methods. 
\par
From a practical point of view,
the nonlinear operator problem is a quite important subject 
to be tackled.
In many problems of chemical interest,
the real time correlation functions of nonlinear operators
are often 
needed for the calculation of various dynamical properties \cite{McQ}.
For example, 
the CMD has been applied to the calculation of 
the transport coefficients such as
thermal conductivity, shear viscosity, and bulk viscosity
of quantum liquid parahydrogen \cite{appli}.
Here it is found that 
the calculated transport properties 
are in good agreement with the experimental data.
However, there is no rigorous theoretical basis for applying 
the CMD method to such properties
represented by means of 
the correlation functions involving
operators nonlinear in $\hat{q}$ (or momentum operator $\hat{p}$). 
Therefore we need, in general, a quantum dynamics method 
which is theoretically valid even 
for the time correlation functions of nonlinear operators.

\par
The present status of the nonlinear operator problem
in the three quantum dynamics methods, CMD, RPMD, and EPAC,
is summarized as follows.
For the CMD method, 
Reichman {\it et al.} have argued this problem 
in their pioneering paper
to conclude that 
a CMD correlation function involving nonlinear operators
corresponds to 
a higher-order Kubo-transformed correlation function \cite{reich}.
On the other hand, Craig and Manolopoulos have shown that 
the RPMD is exact
for all the operators involving $\hat{q}$ 
in the limit $t\to 0$ \cite{cm}.
As for our EPAC method, 
the nonlinear operator problem
has not been examined yet.
In addition to these three methods, 
a theoretical approach based on the quantum mode-coupling theory 
has been applied to study 
the dynamical properties involving nonlinear operators 
in quantum liquids \cite{rr1,rr2}. 
\par
In the present paper, we develop the method of the EPAC
for the calculation of the real time quantum correlation function
involving the nonlinear operator $\hat{A}=\hat{B}=\hat{q}^{n}$.
As a simple example,
at first we show the EPAC correlation function 
$\langle \hat{q}^{2}(t)\hat{q}^{2}(0)\rangle_{\beta}^{EPAC}$
for a harmonic oscillator comparing with the results of 
the CMD and the RPMD. 
Next we calculate 
$\langle \hat{q}^{2}(t)\hat{q}^{2}(0)\rangle_{\beta}^{EPAC}$
numerically in an asymmetric anharmonic system.
We also propose a simplified EPAC method
to reduce the computational cost required in the EPAC calculation.
\par
In Sec. II, we summarize the effective action formalism 
and present how to calculate the EPAC correlation function involving 
nonlinear operators. 
The results for a harmonic oscillator are given in Sec. III.
Numerical results for an anharmonic oscillator are shown in Sec. IV.
In this section we also present the simplified EPAC method.
The conclusions are given in Sec. V.
\par
\section{EFFECTIVE POTENTIAL ANALYTIC CONTINUATION METHOD 
FOR NONLINEAR OPERATORS}
\hspace*{\parindent}
Hereafter we treat 
the real time quantum autocorrelation function of the nonlinear 
operator $\hat{q}^{n}$,
$~\langle \hat{q}^{n}(t)\hat{q}^{n}(0)\rangle_{\beta}$.
In principle, this can be obtained
from the imaginary time Green function 
$\langle T \hat{q}^{n}(\tau)\hat{q}^{n}(0)\rangle_{\beta}$
via an analytic continuation procedure \cite{bell}.
Here $T$ represents a time-ordered product.
On the other hand, it is known that the Green function 
$\langle T \hat{q}^{n}(\tau)\hat{q}^{n}(0)\rangle_{\beta}$
is given as a special case of $2n$-point imaginary time Green function
$\langle T \hat{q}(\tau_{1})\cdot\cdot\cdot\hat{q}(\tau_{2n})\rangle_{\beta}$,
which can be constructed from the standard effective potential $V_{\beta}(Q)$
appearing in the effective action formalism \cite{ep,riv,ps}.  
Consequently, the real time quantum correlation function 
$\langle \hat{q}^{n}(t)\hat{q}^{n}(0)\rangle_{\beta}$
should be obtained by means of 
the effective action formalism and the analytic continuation.
A series of these procedures is the EPAC method 
for the nonlinear operator $\hat{q}^{n}$,
which we newly show in this section.
As the simplest example,
we present the EPAC calculation of
the autocorrelation function of the quadratic operator $\hat{q}^{2}$,  
$\langle \hat{q}^{2}(t)\hat{q}^{2}(0)\rangle_{\beta}$,
in Secs. IIA and IIB.

\subsection{Effective action formalism for multipoint Green functions}
\hspace*{\parindent}
We begin with
the effective action formalism \cite{ep,riv,ps}.
Consider a quantum system where a quantum particle of mass $m$ moves 
in a one-dimensional potential $V(q)$ 
at inverse temperature $\beta$.
The quantum canonical partition function of this system 
is expressed
in terms of the imaginary time path integral
\begin{eqnarray}
{\cal Z}_{\beta}[J]&=&\int^{\infty}_{-\infty}dq
\int^{q(\beta\hbar)=q}_{q(0)=q}{\cal D}q
~\!\exp\left[
-S_{E}/\hbar
+\int^{\beta\hbar}_{0}d\tau J(\tau)q(\tau)/\hbar
\right]
,\label{2}
\end{eqnarray}
where $S_{E}$ is the
Euclidean action functional
\begin{eqnarray}
S_{E}[q]=\int^{\beta\hbar}_{0}d\tau
\left[~\!\frac{1}{2}~\!m ~\!\dot{q}^2 + V(q)~\!\right]
, \label{3}
\end{eqnarray}
and $J(\tau)$ is an external source.
The generating functional  $W_{\beta}[J]$ is defined as 
\begin{eqnarray}
W_{\beta}[J]
&=& \hbar \log {\cal Z}_{\beta}[J] 
.\label{4}
\end{eqnarray}
The functional derivative of $W_{\beta}[J]$ with respect to $J(\tau)$ 
produces the quantum statistical-mechanical expectation value of 
the operator $\hat{q}$ in the presence of $J$,
\begin{eqnarray}
\frac{\delta W_{\beta}[J]}{\delta J(\tau)}
&=&\langle \hat{q}(\tau)\rangle_{\beta}^{J} \equiv Q(\tau) 
.\label{5}
\end{eqnarray}
The effective action $\Gamma_{\beta}[Q]$ is defined 
by the Legendre transform of $W_{\beta}[J]$,
\begin{eqnarray}
\Gamma_{\beta}[Q]&=&-W_{\beta}[J]
+\int^{\beta\hbar}_{0}d\tau J(\tau)Q(\tau)
,\label{6}
\end{eqnarray}
which satisfies the quantum-mechanical Euler-Lagrange equation
$\delta\Gamma_{\beta}[Q]/\delta Q(\tau)=J(\tau)$.
The exact quantum statistical-mechanical expectation value 
$Q_{S}(\tau)$($=\langle \hat{q}(\tau)\rangle_{\beta}$) 
is obtained as a solution of the equation 
\begin{eqnarray}
\frac{\delta\Gamma_{\beta}[Q]}{\delta Q(\tau)}=0
.\label{8}
\end{eqnarray}
Note that the expectation value $Q_{S}(\tau)$
is independent of imaginary time $\tau$ in thermal equilibrium.
\par
The $n$-point connected Green function 
$G_{\beta}^{C}(\tau_{1},...,\tau_{n})$
is generated by the $n$th-order functional derivative of $W_{\beta}[J]$ 
with respect to $J(\tau)$ \cite{riv,ps}.
For example, the two-point connected Green function is given by
\begin{eqnarray}
G_{\beta}^{C}(\tau_{1}-\tau_{2}) 
&=&
\hbar\left.
\frac{\delta^{2} W_{\beta}[J]}{\delta J(\tau_{1})\delta J(\tau_{2})}
\right|_{J=0}
\nonumber \\
&=&
\langle T \hat{q}(\tau_{1})\hat{q}(\tau_{2})\rangle_{\beta}
-\langle \hat{q}(\tau_{1})\rangle_{\beta}
~\!\langle \hat{q}(\tau_{2})\rangle_{\beta}
.\label{9}
\end{eqnarray}
Then we obtain the two-point Green function
\begin{eqnarray}
\langle T \hat{q}(\tau_{1})\hat{q}(\tau_{2})\rangle_{\beta}
=G_{\beta}^{C}(\tau_{1}-\tau_{2})
+Q_{S}(\tau_{1})Q_{S}(\tau_{2})
.\label{10}
\end{eqnarray}
The procedure in Eqs. (\ref{2})-(\ref{10})
has been described in Ref. \onlinecite{hk1}. 
In a similar way, the three- and four-point Green functions can be 
explicitly expressed as
\begin{eqnarray}
\langle T \hat{q}(\tau_{1})\hat{q}(\tau_{2})\hat{q}(\tau_{3})\rangle_{\beta}
&=&
\hbar^{2}\left.\frac{\delta^{3} W_{\beta}[J]}
{\delta J(\tau_{1})\delta J(\tau_{2})\delta J(\tau_{3})}
\right|_{J=0}
\nonumber \\
&&+Q_{S}(\tau_{1})
\langle T \hat{q}(\tau_{2})\hat{q}(\tau_{3})\rangle_{\beta}
+(1\leftrightarrow 2)+(1\leftrightarrow 3)
\nonumber \\
&&-~2~Q_{S}(\tau_{1})Q_{S}(\tau_{2})Q_{S}(\tau_{3})
,\\ \label{11}
\langle T \hat{q}(\tau_{1})\hat{q}(\tau_{2})\hat{q}(\tau_{3})
\hat{q}(\tau_{4})\rangle_{\beta}
&=&
\hbar^{3}\left.\frac{\delta^{4} W_{\beta}[J]}
{\delta J(\tau_{1})\delta J(\tau_{2})\delta J(\tau_{3})\delta J(\tau_{4})}
\right|_{J=0}
\nonumber \\
&&\!\!\!\!\!\!\!\!\!\!\!\!\!\!\!
+\hbar^{2}Q_{S}(\tau_{1})\left.\frac{\delta^{3} W_{\beta}[J]}
{\delta J(\tau_{2})\delta J(\tau_{3})\delta J(\tau_{4})}
\right|_{J=0}
+(1\leftrightarrow 2)+(1\leftrightarrow 3)+(1\leftrightarrow 4)
\nonumber \\
&&\!\!\!\!\!\!\!\!\!\!\!\!\!\!\!
+\langle T \hat{q}(\tau_{1})\hat{q}(\tau_{2})\rangle_{\beta}
\langle T \hat{q}(\tau_{3})\hat{q}(\tau_{4})\rangle_{\beta}
+(2\leftrightarrow 3)+(2\leftrightarrow 4)
\nonumber \\
&&\!\!\!\!\!\!\!\!\!\!\!\!\!\!\!
-~2~Q_{S}(\tau_{1})Q_{S}(\tau_{2})Q_{S}(\tau_{3})Q_{S}(\tau_{4})
,\label{12}
\end{eqnarray}
respectively.
Here $(i\leftrightarrow j)$ denotes the interchange 
of $\tau_{i}$ with $\tau_{j}$.
\par
On the other hand, 
the $n$th-order functional derivative 
$\delta^{n}W_{\beta}/\delta J^{n}$ is connected with 
the $n$th-order functional derivative of 
the effective action $\delta^{n}\Gamma_{\beta}/\delta Q^{n}$ \cite{ps},
\begin{eqnarray}
\left.
\frac{\delta^{2} W_{\beta}[J]}{\delta J(\tau_{1})\delta J(\tau_{2})}
\right|_{J=0}
&=&
\left(\left.
\frac{\delta^{2} \Gamma_{\beta}[Q]}{\delta Q(\tau_{1})\delta Q(\tau_{2})}
\right|_{Q=Q_{S}}\right)^{-1}
,\label{13} \\ 
\left.
\frac{\delta^{3} W_{\beta}[J]}
{\delta J(\tau_{1})\delta J(\tau_{2})\delta J(\tau_{3})}
\right|_{J=0}
&=&
-\frac{1}{\hbar^{3}}
\int^{\beta\hbar}_{0}\!\!\!\!\!ds_{1}
\int^{\beta\hbar}_{0}\!\!\!\!\!ds_{2}
\int^{\beta\hbar}_{0}\!\!\!\!\!ds_{3} 
\nonumber \\ 
&&\times G_{\beta}^{C}(s_{1}-\tau_{1})
G_{\beta}^{C}(s_{2}-\tau_{2})
G_{\beta}^{C}(s_{3}-\tau_{3})
\nonumber \\
&&\times\left.
\frac{\delta^{3} \Gamma_{\beta}[Q]}
{\delta Q(s_{1})\delta Q(s_{2})\delta Q(s_{3})}
\right|_{Q=Q_{S}}
,\label{14}\\ 
\left.
\frac{\delta^{4} W_{\beta}[J]}
{\delta J(\tau_{1})\delta J(\tau_{2})\delta J(\tau_{3})\delta J(\tau_{4})}
\right|_{J=0}
&=&
-\frac{1}{\hbar^{4}}
\int^{\beta\hbar}_{0}\!\!\!\!\!ds_{1}
\int^{\beta\hbar}_{0}\!\!\!\!\!ds_{2}
\int^{\beta\hbar}_{0}\!\!\!\!\!ds_{3}
\int^{\beta\hbar}_{0}\!\!\!\!\!ds_{4}
\nonumber \\
&&\times G_{\beta}^{C}(s_{1}-\tau_{1})
G_{\beta}^{C}(s_{2}-\tau_{2})
G_{\beta}^{C}(s_{3}-\tau_{3})
G_{\beta}^{C}(s_{4}-\tau_{4})
\nonumber \\
&&\times\left.
\frac{\delta^{4} \Gamma_{\beta}[Q]}
{\delta Q(s_{1})\delta Q(s_{2})\delta Q(s_{3})\delta Q(s_{4})}
\right|_{Q=Q_{S}}
\nonumber \\
&&\!\!\!\!\!\!\!\!\!\!\!\!\!\!\!\!\!\!\!\!
+\frac{1}{\hbar^{5}}
\int^{\beta\hbar}_{0}\!\!\!\!\!ds_{1}
\int^{\beta\hbar}_{0}\!\!\!\!\!ds_{2}
\int^{\beta\hbar}_{0}\!\!\!\!\!ds_{3}
\int^{\beta\hbar}_{0}\!\!\!\!\!ds_{4}
\int^{\beta\hbar}_{0}\!\!\!\!\!du
\int^{\beta\hbar}_{0}\!\!\!\!\!dv
\nonumber \\
&&\!\!\!\!\!\!\!\!\!\!\!\!\!\!\!\!\!\!\!\!
\times G_{\beta}^{C}(s_{1}-\tau_{1})
G_{\beta}^{C}(s_{2}-\tau_{2})
G_{\beta}^{C}(u-v)
G_{\beta}^{C}(s_{3}-\tau_{3})
G_{\beta}^{C}(s_{4}-\tau_{4})
\nonumber \\
&&\!\!\!\!\!\!\!\!\!\!\!\!\!\!\!\!\!\!\!\!
\times\left.
\frac{\delta^{3} \Gamma_{\beta}[Q]}
{\delta Q(s_{1})\delta Q(s_{2})\delta Q(u)}
\right|_{Q=Q_{S}}
\!\!\!\!\cdot~\left.
\frac{\delta^{3} \Gamma_{\beta}[Q]}
{\delta Q(v)\delta Q(s_{3})\delta Q(s_{4})}
\right|_{Q=Q_{S}}
\nonumber \\
&&\!\!\!\!\!\!\!\!\!\!\!\!\!\!\!\!\!\!\!\!
+(2\leftrightarrow 3)+(2\leftrightarrow 4)
.\label{15}
\end{eqnarray}
Using Eqs. (\ref{9})-(\ref{15}), we can construct 
the imaginary time Green functions in terms of 
the effective action $\Gamma_{\beta}[Q]$.\par
Now we employ the local potential approximation (LPA) 
to the effective action,
\begin{eqnarray}
\Gamma_{\beta}[Q]&=&\int^{\beta\hbar}_{0}d\tau\left[
V_{\beta}(Q)+\frac{1}{2}~\!m~\!\dot{Q}^{2}
\right]
,\label{16}
\end{eqnarray}
where $V_{\beta}(Q)$ is the standard effective potential, 
i.e., the leading order of the derivative expansion 
of $\Gamma_{\beta}[Q]$ \cite{ep,riv,ps}. 
For the case $Q_{S}(\tau)$ is $\tau$ independent, 
Eq. (\ref{8}) becomes the stationary condition
$\left.{\partial V_{\beta}}/{\partial Q} 
\right|_{Q=Q_{\rm min}}=0$,
which determines
the expectation value $Q_{S}(\tau)$
as the standard effective potential minimum $Q_{\rm min}$.
Then the functional derivatives of 
the effective action $\delta^{n}\Gamma_{\beta}/\delta Q^{n}$ become
\begin{eqnarray}
\left.
\frac{\delta^{2} \Gamma_{\beta}[Q]}{\delta Q(\tau_{1})\delta Q(\tau_{2})}
\right|_{Q=Q_{S}}
&=&
\left(-m\frac{d^{2}}{d\tau^{2}}+
\left.
\frac{\partial^{2} V_{\beta}}{\partial Q^{2}}
\right|_{Q=Q_{\rm min}}
\right)\delta(\tau_{1}-\tau_{2}),
\label{17}\\ 
\left.
\frac{\delta^{3} \Gamma_{\beta}[Q]}
{\delta Q(\tau_{1})\delta Q(\tau_{2})\delta Q(\tau_{3})}
\right|_{Q=Q_{S}}
&=&
\left.
\frac{\partial^{3} V_{\beta}}{\partial Q^{3}}
\right|_{Q=Q_{\rm min}}
\delta(\tau_{1}-\tau_{2})\delta(\tau_{1}-\tau_{3}),
\label{18}\\ 
\left.
\frac{\delta^{4} \Gamma_{\beta}[Q]}
{\delta Q(\tau_{1})\delta Q(\tau_{2})\delta Q(\tau_{3})\delta Q(\tau_{4})}
\right|_{Q=Q_{S}}
&=&
\left.
\frac{\partial^{4} V_{\beta}}{\partial Q^{4}}
\right|_{Q=Q_{\rm min}}
\delta(\tau_{1}-\tau_{2})\delta(\tau_{1}-\tau_{3})\delta(\tau_{1}-\tau_{4})
.\label{19}
\end{eqnarray}
If we expand $V_{\beta}$ around the minimum $Q=Q_{\rm min}$,
\begin{eqnarray}
V_{\beta}(Q)=\sum_{n=0}^{\infty}
\frac{a_{n}}{n!}(Q-Q_{\rm min})^{n}
,\label{20}
\end{eqnarray}
then the derivatives of $V_{\beta}$ 
appearing in Eqs. (\ref{17})-(\ref{19})
are given as
the coefficients in the series, 
$a_{n}=\partial^{n}V_{\beta}/\partial Q^{n}|_{Q=Q_{\rm min}}$.

\subsection{EPAC method}
\hspace*{\parindent}
Now we proceed to the calculations of    
the real time quantum correlation function
$\langle \hat{q}^{2}(t)\hat{q}^{2}(0)\rangle_{\beta}$,
i.e., the aim of the present paper.
First, by setting 
$\tau_{1}=\tau_{2}=\tau ~(0\le\tau\le\beta\hbar)$ 
and $\tau_{3}=\tau_{4}=0$ in Eq. (\ref{12}),
we obtain the imaginary time Green function 
\begin{eqnarray}
\langle T \hat{q}^{2}(\tau)\hat{q}^{2}(0)\rangle_{\beta}
&=&
\hbar^{3}\left.\frac{\delta^{4} W_{\beta}[J]}
{\delta J(\tau)\delta J(\tau)\delta J(0)\delta J(0)}
\right|_{J=0}
\nonumber \\
&+&
2\hbar^{2}Q_{\rm min}\left[
\left.\frac{\delta^{3} W_{\beta}[J]}
{\delta J(\tau)\delta J(\tau)\delta J(0)}
\right|_{J=0}
+
\left.\frac{\delta^{3} W_{\beta}[J]}
{\delta J(\tau)\delta J(0)\delta J(0)}
\right|_{J=0}
\right]
\nonumber \\
&+&
2\left(\langle T \hat{q}(\tau)\hat{q}(0)\rangle_{\beta}\right)^{2}
+\left(\langle T \hat{q}(0)\hat{q}(0)\rangle_{\beta}\right)^{2}
-~2~Q_{\rm min}^{4}
.\label{21}
\end{eqnarray}
With the definition 
$\omega_{\beta}=\sqrt{a_{2}/m}$ and 
$\alpha=\beta\hbar\omega_{\beta}/2$,
from Eqs.  (\ref{9}), (\ref{10}), 
(\ref{13})-(\ref{15}), and (\ref{17})-(\ref{20}),
the components of Eq. (\ref{21}) can approximately be written as 
\begin{eqnarray}
\langle T \hat{q}(\tau)\hat{q}(0)\rangle_{\beta}
&=&
\frac{\hbar}{2m\omega_{\beta}}
\frac{1}
{e^{\alpha}-e^{-\alpha}}
\left[e^{\alpha}e^{-\omega_{\beta}\tau}+e^{-\alpha}e^{\omega_{\beta}\tau}
\right]+Q_{\rm min}^{2}
,\label{22}
\end{eqnarray}
\begin{eqnarray}
\left.\frac{\delta^{3} W_{\beta}[J]}
{\delta J(\tau)\delta J(\tau)\delta J(0)}
\right|_{J=0}
&+&
\left.\frac{\delta^{3} W_{\beta}[J]}
{\delta J(\tau)\delta J(0)\delta J(0)}
\right|_{J=0}\nonumber \\
&&\!\!\!\!\!\!\!\!\!\!\!\!\!\!\!\!\!\!\!\!\!\!\!\!
=-\frac{a_{3}}{6m^{3}\omega_{\beta}^{4}}
\frac{1}
{\left(e^{\alpha}-e^{-\alpha}\right)^{3}}
\left[
(e^{\alpha}-e^{3\alpha})e^{-2\omega_{\beta}\tau}
+(e^{-3\alpha}-e^{-\alpha})e^{2\omega_{\beta}\tau}
\right.\nonumber \\
&&\!\!\!\!\!\!\!\!\!\!\!\!
+\left.2(e^{3\alpha}-e^{-\alpha})e^{-\omega_{\beta}\tau}
+2(e^{\alpha}-e^{-3\alpha})e^{\omega_{\beta}\tau}
+6(e^{\alpha}-e^{-\alpha})
\right]
,\label{23}
\end{eqnarray}
\begin{eqnarray}
&&\left.\frac{\delta^{4} W_{\beta}[J]}
{\delta J(\tau)\delta J(\tau) \delta J(0)\delta J(0)}
\right|_{J=0}
\nonumber \\
&=&
-\frac{a_{4}}{32m^{4}\omega_{\beta}^{5}}
\frac{1}
{\left(e^{\alpha}-e^{-\alpha}\right)^{4}}
\left[~
4\alpha
(4+e^{-2\omega_{\beta}\tau}+e^{2\omega_{\beta}\tau})\right.
\nonumber \\
&&
+2(e^{4\alpha}-1)\omega_{\beta}\tau e^{-2\omega_{\beta}\tau}
+2(e^{-4\alpha}-1)\omega_{\beta}\tau e^{2\omega_{\beta}\tau}
\nonumber \\
&&\left.
+8(e^{2\alpha}-e^{-2\alpha})
+(e^{4\alpha}-1)e^{-2\omega_{\beta}\tau}
+(1-e^{-4\alpha})e^{2\omega_{\beta}\tau}
~\right]
\nonumber \\
&&
+\frac{a_{3}^{2}}{288m^{5}\omega_{\beta}^{7}}
\frac{1}
{\left(e^{\alpha}-e^{-\alpha}\right)^{5}}
\left[~
60\alpha
(e^{\alpha}-e^{-\alpha})
(4+e^{-2\omega_{\beta}\tau}+e^{2\omega_{\beta}\tau})\right.
\nonumber \\
&&
+30(e^{5\alpha}-e^{3\alpha}-e^{\alpha}+e^{-\alpha})
\omega_{\beta}\tau e^{-2\omega_{\beta}\tau}
+30(-e^{\alpha}+e^{-\alpha}+e^{-3\alpha}-e^{-5\alpha})
\omega_{\beta}\tau e^{2\omega_{\beta}\tau}
\nonumber \\
&&
+16(e^{5\alpha}-2e^{3\alpha}+e^{\alpha})e^{-3\omega_{\beta}\tau}
+16(e^{-\alpha}-2e^{-3\alpha}+e^{-5\alpha})e^{3\omega_{\beta}\tau}
\nonumber \\
&&
-17(e^{5\alpha}-e^{3\alpha}-e^{\alpha}+e^{-\alpha})e^{-2\omega_{\beta}\tau}
-17(e^{\alpha}-e^{-\alpha}-e^{-3\alpha}+e^{-5\alpha})e^{2\omega_{\beta}\tau}
\nonumber \\
&&
+16(e^{5\alpha}+3e^{3\alpha}-8e^{\alpha}+3e^{-\alpha}+e^{-3\alpha})
e^{-\omega_{\beta}\tau}
+16(e^{3\alpha}+3e^{\alpha}-8e^{-\alpha}+3e^{-3\alpha}+e^{-5\alpha})
e^{\omega_{\beta}\tau}
\nonumber \\
&&\left.
+248(e^{3\alpha}-e^{\alpha}-e^{-\alpha}+e^{-3\alpha})
~\right].\label{24}
\end{eqnarray}
Substituting Eqs. (\ref{22})-(\ref{24}) into Eq. (\ref{21}), 
we can approximately represent
the quadratic imaginary time Green function 
 $\langle T \hat{q}^{2}(\tau)\hat{q}^{2}(0)\rangle_{\beta}$
in terms of the quantities 
$Q_{\rm min}$, $\omega_{\beta}$, $a_{3}$, and $a_{4}$.
\par
Then, by means of the analytic continuation \cite{bell},
the EPAC correlation function 
$\langle \hat{q}^{2}(t)\hat{q}^{2}(0)\rangle_{\beta}^{\rm EPAC}$
is obtained from the imaginary time quantity 
$\langle T \hat{q}^{2}(\tau)\hat{q}^{2}(0)\rangle_{\beta}$. 
This is the final step of the EPAC method
for the correlation function of the nonlinear operator $\hat{q}^{2}$.
The EPAC correlation function,
which is the approximation to the exact correlation function 
$\langle \hat{q}^{2}(t)\hat{q}^{2}(0)\rangle_{\beta}$,
can be expressed in a simple form
\begin{eqnarray}
\langle \hat{q}^{2}(t)\hat{q}^{2}(0)\rangle_{\beta}^{\rm EPAC}
=a_{4}A(t)+a_{3}^{2}B(t)+a_{3}Q_{\rm min}C(t)+D(t),
\label{25}
\end{eqnarray}
where
\begin{eqnarray}
A(t)&=&
-\frac{\hbar^{3}}{32m^{4}\omega_{\beta}^{5}}
\frac{1}
{(e^{\alpha}-e^{-\alpha})^{4}}
\left[~
8(e^{2\alpha}-e^{-2\alpha})
+8\alpha(\cos 2\omega_{\beta}t+2)\right.
\nonumber \\
&&+(e^{4\alpha}-e^{-4\alpha})
(2\omega_{\beta}t\sin 2\omega_{\beta}t+\cos 2\omega_{\beta}t)
\nonumber \\
&&\left.
+~i~(e^{4\alpha}+e^{-4\alpha}-2)
(2\omega_{\beta}t\cos 2\omega_{\beta}t-\sin 2\omega_{\beta}t)~\right],
\label{25a}\\
B(t)&=&
\frac{\hbar^{3}}{288m^{5}\omega_{\beta}^{7}}
\frac{1}
{(e^{\alpha}-e^{-\alpha})^{5}}
\left[~
248(e^{3\alpha}-e^{\alpha}-e^{-\alpha}+e^{-3\alpha})
+120\alpha(e^{\alpha}-e^{-\alpha})(\cos 2\omega_{\beta}t+2)\right.
\nonumber \\
&&+16
(e^{5\alpha}-2e^{3\alpha}+e^{\alpha}+e^{-\alpha}-2e^{-3\alpha}+e^{-5\alpha})
\cos 3\omega_{\beta}t
\nonumber \\
&&+(e^{5\alpha}-e^{3\alpha}-e^{-3\alpha}+e^{-5\alpha})
(30\omega_{\beta}t\sin 2\omega_{\beta}t-17\cos 2\omega_{\beta}t)
\nonumber \\
&&+16(e^{5\alpha}+4e^{3\alpha}-5e^{\alpha}-5e^{-\alpha}
+4e^{-3\alpha}+e^{-5\alpha})
\cos \omega_{\beta}t
\nonumber \\
&&+~i~\left\{
-16
(e^{5\alpha}-2e^{3\alpha}+e^{\alpha}-e^{-\alpha}+2e^{-3\alpha}-e^{-5\alpha})
\sin 3\omega_{\beta}t \right.
\nonumber \\
&&~~~~~~~
+(e^{5\alpha}-e^{3\alpha}-2e^{\alpha}+2e^{-\alpha}+e^{-3\alpha}-e^{-5\alpha})
(30\omega_{\beta}t\cos 2\omega_{\beta}t+17\sin 2\omega_{\beta}t)
\nonumber \\
&&~~~~~~~\left.\left.
-16(e^{5\alpha}+2e^{3\alpha}-11e^{\alpha}+11e^{-\alpha}
-2e^{-3\alpha}-e^{-5\alpha})
\sin \omega_{\beta}t~\right\}~\right],
\label{25b}\\
C(t)&=&
-\frac{\hbar^{2}}{3m^{3}\omega_{\beta}^{4}}
\frac{1}
{(e^{\alpha}-e^{-\alpha})^{3}}
\left[~
6(e^{\alpha}-e^{-\alpha})
\right.
\nonumber \\
&&
-(e^{3\alpha}-e^{\alpha}+e^{-\alpha}-e^{-3\alpha})\cos 2\omega_{\beta}t
+2(e^{3\alpha}+e^{\alpha}-e^{-\alpha}-e^{-3\alpha})\cos \omega_{\beta}t
\nonumber \\
&&\left.
+~i~(e^{3\alpha}-e^{\alpha}-e^{-\alpha}+e^{-3\alpha})
(\sin 2\omega_{\beta}t-2\sin \omega_{\beta}t)~\right],
\label{25c}\\
D(t)&=&
\frac{\hbar^2}{4m^{2}\omega_{\beta}^{2}}\left[~
2\coth \alpha
(\coth 2\alpha\cos 2\omega_{\beta} t - i\sin 2\omega_{\beta} t)
+2\coth^2\alpha-1
~\right]
\nonumber \\
&&+\frac{\hbar Q_{\rm min}^{2}}{m\omega_{\beta}}
\left[~2\coth\alpha\cos\omega_{\beta} t
-i~2\sin\omega_{\beta} t +\coth\alpha
~\right]+Q_{\rm min}^{4}.
\label{25d}
\end{eqnarray}
It should be noted that this correlation function, Eq. (\ref{25}), 
consists of three oscillation modes with the frequencies 
$\omega_{\beta}$, $2\omega_{\beta}$, and $3\omega_{\beta}$;
the coefficients $a_{3}$ and $a_{4}$ contribute 
only to the amplitude of the oscillations.
\par
Similarly, the higher-order real time correlation functions 
$\langle \hat{q}^{n}(t)\hat{q}^{n}(0)\rangle_{\beta}$
can approximately be obtained as 
$\langle \hat{q}^{n}(t)\hat{q}^{n}(0)\rangle_{\beta}^{\rm EPAC}$
using the information of the standard effective potential
$V_{\beta}(Q)$.

\section{SIMPLE EXAMPLE : A HARMONIC SYSTEM}
In this section, as a simple example we show the calculation of 
the EPAC correlation function
$\langle \hat{q}^{2}(t)\hat{q}^{2}(0)\rangle_{\beta}^{\rm EPAC}$
for a quantum harmonic oscillator whose classical potential is given by
\begin{eqnarray}
V(q)=\frac{1}{2}~\!m\omega^{2}q^{2}
.\label{26}
\end{eqnarray}
Next, the results of 
the other quantum dynamics methods, 
the CMD \cite{cv} and the RPMD \cite{cm},
are shown. 
\subsection{EPAC correlation function for a harmonic oscillator}
\hspace*{\parindent}
The standard effective potential for the system (\ref{26}) 
is obtained as \cite{hk2} 
\begin{eqnarray}
V_{\beta}(Q)
&=&\frac{1}{2}m\omega^{2}Q^{2}+
\frac{1}{\beta}\log
\left(2\sinh\frac{\beta\hbar\omega}{2}\right).
\label{27}
\end{eqnarray}
For the harmonic system, the minimum of $V_{\beta}(Q)$ is located
at the point $Q=Q_{\rm min}=0$, while 
the effective frequency is 
$\omega_{\beta}=\omega$.
Furthermore, it is evident that 
the higher-order derivatives of $V_{\beta}(Q)$
vanish, $a_{3}=a_{4}=0$. 
Consequently, the EPAC correlation function (\ref{25})
becomes 
\begin{eqnarray}
\!\!\!\!\!\!\!\!\!\!
\langle \hat{q}^{2}(t)\hat{q}^{2}(0)\rangle_{\beta}^{\rm EPAC}
=
\frac{\hbar^2}{4m^{2}\omega^{2}}\!\left[
2\coth \frac{\beta\hbar\omega}{2}
\left(\coth \beta\hbar\omega
\cos 2\omega t - i\sin 2\omega t
\right)
+2\coth^2\frac{\beta\hbar\omega}{2}
-1\right],
\label{28}
\end{eqnarray}
which is equal to the exact quantum correlation function
$\langle \hat{q}^{2}(t)\hat{q}^{2}(0)\rangle_{\beta}$.
That is, the EPAC method is exact in the harmonic system (\ref{26})
at any temperature. 
This is because the LPA [Eq. (\ref{16})], which is  
the only approximation employed in the EPAC method,
is exact for harmonic systems \cite{nprg}.
Therefore, it can be shown that 
the EPAC correlation function 
$\langle \hat{q}^{n}(t)\hat{q}^{n}(0)\rangle_{\beta}^{\rm EPAC}$
is exact for any $n$ for harmonic systems.

\subsection{The other quantum dynamics methods}
\hspace*{\parindent}
In harmonic systems, both the CMD and the RPMD
are exact for the linear operator \cite{jv,cm}.
That is, both 
the CMD correlation function and the RPMD correlation function
are equal to the exact canonical, or Kubo-transformed, 
correlation function
$\langle \hat{q}(t)\hat{q}(0)\rangle_{\beta}^{\rm can}
=(1/\beta)\int_{0}^{\beta}d\lambda
~\langle \hat{q}(t-i\hbar\lambda)\hat{q}(0)\rangle_{\beta}$.
However, even in harmonic systems, these methods are no longer exact 
for the correlation functions of nonlinear operators \cite{jv,cm}.
\par
In the harmonic system (\ref{26}),
we explicitly present  
the exact canonical autocorrelation function of
the nonlinear operator $\hat{q}^{2}$  
\begin{eqnarray}
\langle \hat{q}^{2}(t)\hat{q}^{2}(0)\rangle_{\beta}^{\rm can}
&=&\frac{\hbar^2}{4m^{2}\omega^{2}}\left[
\frac{2}{\beta\hbar\omega}
\coth\frac{\beta\hbar\omega}{2}
\cos 2\omega t
+2\coth^2\frac{\beta\hbar\omega}{2}-1\right]
,\label{29}
\end{eqnarray}
while the corresponding 
CMD correlation function with the {\it classical operator} \cite{cv}
is obtained as
\begin{eqnarray}
\langle q^{2}_{c}(t) q^{2}_{c}(0)\rangle_{\beta}^{\rm CMD}
&=&\frac{1}{\beta^{2}m^{2}\omega^{4}}(\cos 2\omega t +2)
,\label{30}
\end{eqnarray}
where $q_{c}$ is the position centroid variable. 
The CMD correlation function 
with the {\it effective classical operator} \cite{jv}
is also obtained as
\begin{eqnarray}
\langle q^{2}_{c}(t) (q^{2})^{c}_{\beta}(0)\rangle_{\beta}^{\rm CMD}
=\frac{1}{\beta^{2}m^{2}\omega^{4}}
[\cos2\omega t+\frac{\beta\hbar\omega}{2}\coth\frac{\beta\hbar\omega}{2}+1],
\label{31b}
\end{eqnarray}
where $(q^{2})^{c}_{\beta}$ is 
the effective classical operator:
$(q^{2})^{c}_{\beta}=q_{c}^{2}+
(\hbar/2m\omega)(\coth(\beta\hbar\omega/2)-2/\beta\hbar\omega)$.
On the other hand, the RPMD correlation function is given by
\begin{eqnarray}
\langle \hat{q}^{2}(t)\hat{q}^{2}(0)\rangle_{\beta}^{\rm RPMD}
&=&\lim_{P\to\infty}
\frac{1}{\beta^{2}m^{2}}\left[
\sum_{n=1}^{P}\frac{1}{\omega_{n}^{4}}(\cos 2\omega_{n} t +1)+
\sum_{n=1}^{P}\sum_{l=1}^{P}\frac{1}{\omega_{n}^{2}\omega_{l}^{2}}
\right]
,\label{31}
\end{eqnarray}
where 
$\omega_{n}=\sqrt{\omega^{2}+(2k_{P}/m)(1-\cos(2\pi n/P))}$,
$k_{P}=mP^{2}/(\beta^{2}\hbar^{2})$, 
and $P$ is the number of discretization of imaginary time. 
\par
\begin{figure}
\begin{tabular}{c}
\includegraphics[width=130mm]{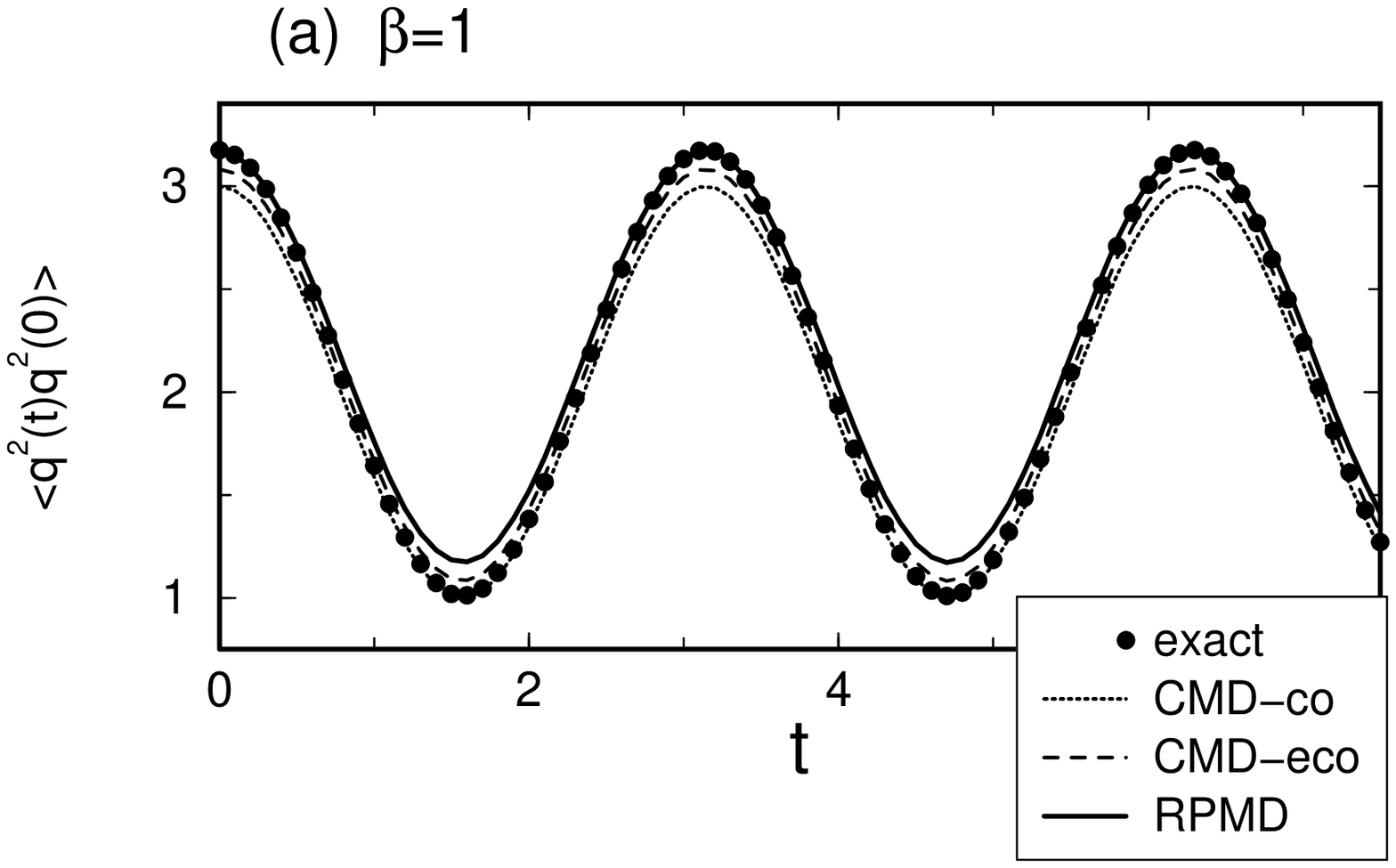}\\
\includegraphics[width=130mm]{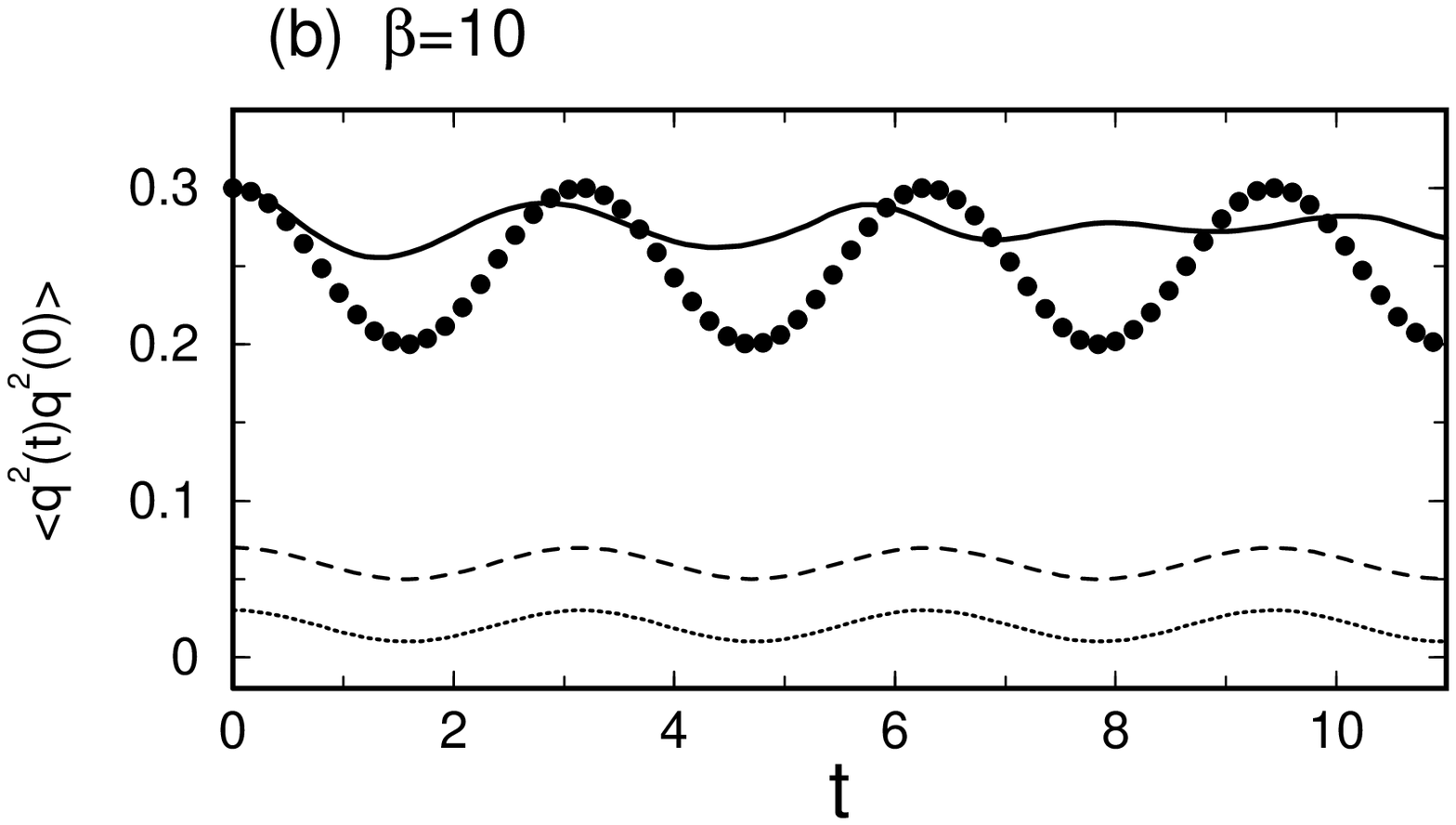}\\
\end{tabular}
\vspace{0mm}
\caption{
The plot of the exact canonical correlation function Eq. (\ref{29}),
the CMD correlation function with the {\it classical operator} 
(CMD-co) Eq. (\ref{30}),
the CMD correlation function with the {\it effective classical operator} 
(CMD-eco) Eq. (\ref{31b}),
and the RPMD correlation function
Eq. (\ref{31})
for the quantum harmonic oscillator (\ref{26}).
(a) at $\beta=1$ and (b) at $\beta=10$. 
}
\label{fig:harm}
\end{figure}

Figure \ref{fig:harm} shows the plot of the four 
correlation functions, Eqs. (\ref{29})-(\ref{31}),
at two different temperatures
with the parameters $\hbar=k_{B}=m=\omega=1$.
Here we have also set $P$ as $1000$, 
which is so large as to make the RPMD results converged sufficiently.  
At higher temperature $\beta=1$, in Fig. \ref{fig:harm} (a),  
all the CMD results and the RPMD result are slightly deviate 
from the exact one.
These deviations become remarkable
at lower temperature $\beta=10$ [Fig. \ref{fig:harm} (b)].
As is observed clearly at $\beta=10$,
the CMD correlation functions [Eqs. (\ref{30}) and (\ref{31b})]
fail to reproduce the exact value at $t=0$ and the exact amplitude,
while they oscillate with the correct frequency $2\omega$.  
On the other hand, 
the RPMD correlation function (\ref{31})
is exact at $t=0$, 
while it damps with time      
because of the dephasing effect caused 
by the mode summation $\sum_{n=1}^{P}$.
This damping is remarkable at lower temperature.
\par
For the CMD method, 
Jang and Voth have explicitly shown that
the CMD should be used for the computation 
of the canonical correlation function
$\langle \hat{A}(t)\hat{B}(0)\rangle_{\beta}^{\rm can}
=(1/\beta)\int_{0}^{\beta}d\lambda
~\langle \hat{A}(t-i\hbar\lambda)\hat{B}(0)\rangle_{\beta}$
where the operator $\hat{A}$ must be linear in 
position and/or momentum operators \cite{jv}.
As for the operator $\hat{B}$, 
the {\it effective classical operator} $B_{\beta}^{c}$ should be used 
for the CMD description of this correlation function \cite{jv}.
The CMD method is known to be problematic when the operator $\hat{A}$
is a nonlinear function.
Afterwards two major approaches have been reported 
to deal with 
this nonlinear operator problem in the CMD method \cite{reich,geva1}.
Here we note these approaches:\par 
(1) The first approach is a theory based on the higher-order Kubo-transformed 
correlation function;
Reichman et al. have pointed out that 
the CMD correlation function with 
the effective classical operator $B_{\beta}^{c}$,
$\langle q^{n}_{c}(t) B^{c}_{\beta}(0)\rangle_{\beta}^{\rm CMD}$,
corresponds to the $n$th-order Kubo-transformed 
correlation function
$\langle \hat{q}^{n}(t)\hat{B}(0)\rangle_{\beta}^{n{\rm th}}$ \cite{reich}.
That is, 
the CMD correlation function with 
effective classical operator,
$\langle q^{2}_{c}(t) (q^{2})^{c}_{\beta}(0)\rangle_{\beta}^{\rm CMD}$,
corresponds to the second-order Kubo-transformed 
correlation function
\begin{eqnarray}
\langle \hat{q}^{2}(t)\hat{q}^{2}(0)\rangle_{\beta}^{\rm 2nd}
=\frac{2}{\beta^{2}}\int_{0}^{\beta}d\lambda
\int_{0}^{\beta}d\eta
~\langle \hat{q}(t-i\hbar\lambda)\hat{q}(t-i\hbar\eta)
\hat{q}^{2}(0)\rangle_{\beta}.
\label{31a}
\end{eqnarray}
For the harmonic oscillator (\ref{26}),
it can readily be shown that 
Eq. (\ref{31b}) is equal to Eq. (\ref{31a}).
Thus, the CMD with effective classical operators 
works very well in this framework.  
However, 
there remains a practical problem 
in converting the higher-order Kubo-transformed 
correlation function 
$\langle \hat{q}^{n}(t)\hat{B}(0)\rangle_{\beta}^{n{\rm th}}$
to the original quantum correlation function 
$\langle \hat{q}^{n}(t)\hat{B}(0)\rangle_{\beta}$;
this conversion would be a complicated procedure
in general.\cite{reich}\par  
(2) The second approach is based on novel expressions for 
physical quantities, which are formulated  
using correlation functions involving linear operators. 
Geva {\it et al.} have shown that 
the quantum reaction rate constant can be expressed 
in terms of the canonical correlation function 
$\langle \hat{q}(t)\hat{B}(0)\rangle_{\beta}^{\rm can}$ \cite{geva1}.
Their approach enables us to evaluate the rate constants via CMD calculations 
without further approximations \cite{geva1,geva2}.

\section{NUMERICAL TESTS : AN ANHARMONIC SYSTEM}
\hspace*{\parindent}
In this section we calculate
the EPAC correlation function
$\langle \hat{q}^{2}(t)\hat{q}^{2}(0)\rangle_{\beta}^{\rm EPAC}$
for an asymmetric anharmonic system
with the classical potential \cite{cv,cm,hk2,jv,reich}
\begin{eqnarray}
V(q)=\frac{1}{2}~\!q^{2}+
\frac{1}{10}~\!q^{3}+
\frac{1}{100}~\!q^{4}
,\label{32}
\end{eqnarray}
where natural units $\hbar=k_{B}=m=1$ are employed.
In the following,
first we show the calculated EPAC correlation functions with the exact
correlation functions 
$\langle \hat{q}^{2}(t)\hat{q}^{2}(0)\rangle_{\beta}$,
and then
we present a simplified version of the EPAC method
to discuss its validity.

\subsection{EPAC results for an anharmonic oscillator}
\begin{figure}
\includegraphics[width=130mm]{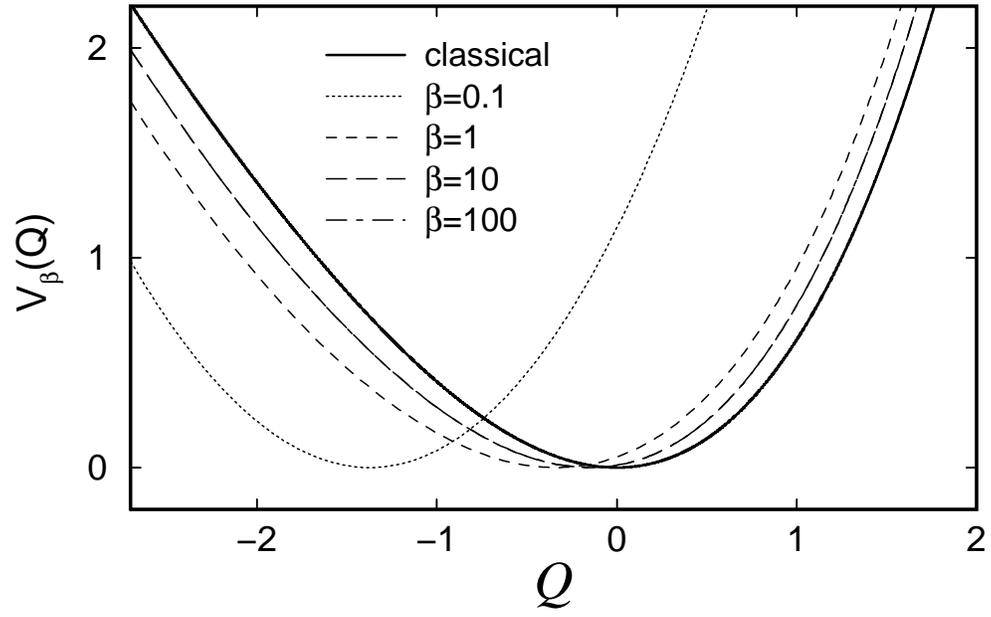}
\vspace{5mm}
\caption{
The inverse temperature $\beta$-dependence of
the standard effective potential $V_{\beta}(Q)$
for the asymmetric anharmonic classical potential (\ref{32}).
In this plot 
we set $V_{\beta}(Q_{\rm min})=0$.
}
\label{fig:ep}
\end{figure}

\hspace*{\parindent}
To evaluate the standard effective potential $V_{\beta}(Q)$,
we need to compute the generating functional $W_{\beta}[J]$ 
[Eq. (\ref{4})]
and to carry out the Legendre transformation (\ref{6}).
There are various computational schemes to evaluate 
the generating functional $W_{\beta}[J]$.
Among them, the PIMD/PIMC technique
is a practical tool to
evaluate $W_{\beta}[J]$ directly or indirectly \cite{hk1,hk2,owy}.
Here we have, however, 
employed the renormalization group (RG) method,
which is suitable for precise calculation of $W_{\beta}[J]$ \cite{drg}. 
Then the standard effective potential $V_{\beta}(Q)$ 
has been computed 
by means of the numerical Legendre transformation \cite{owy}. 
Figure \ref{fig:ep} shows the evaluated standard effective potentials 
at various inverse temperatures $\beta=$ 0.1, 1, 10, and 100.
Minimizing the effective potential $V_{\beta}(Q)$, we have obtained the quantities 
$Q_{\rm min}$, $\omega_{\beta}$, $a_{3}$, and $a_{4}$.
Table \ref{tab:table1} lists the results of computed quantities
for the system (\ref{32}). 
\begin{table}
\caption{\label{tab:table1}
The results of the quantities 
for the quantum anharmonic oscillator (\ref{32}).}
\begin{tabular}{|c|c|c|c|c|}  
\hline 
$\beta$
& $Q_{\rm min}$
& $\omega_{\beta}$
& $a_{3}$
& $a_{4}$ \\
\hline 
~~0.1~  
~~&~~ -1.3735019
~~&~~ 1.07083695
~~&~~ 0.10132291
~~&~~ 0.1018375
~~\\
~~1~  
~~&~~ -0.3375973
~~&~~ 0.91069063
~~&~~ 0.41549732
~~&~~ 0.3305302
~~\\
~~10~  
~~&~~ -0.1501482
~~&~~ 0.96628105
~~&~~ 0.54407872
~~&~~ 0.2606658
~~\\
~~100~
~~&~~ -0.1501276
~~&~~ 0.96631313
~~&~~ 0.54396628
~~&~~ 0.2608735
~~\\
\hline
\end{tabular}

\end{table}
\par
\begin{figure}
\begin{tabular}{c}
\includegraphics[width=130mm]{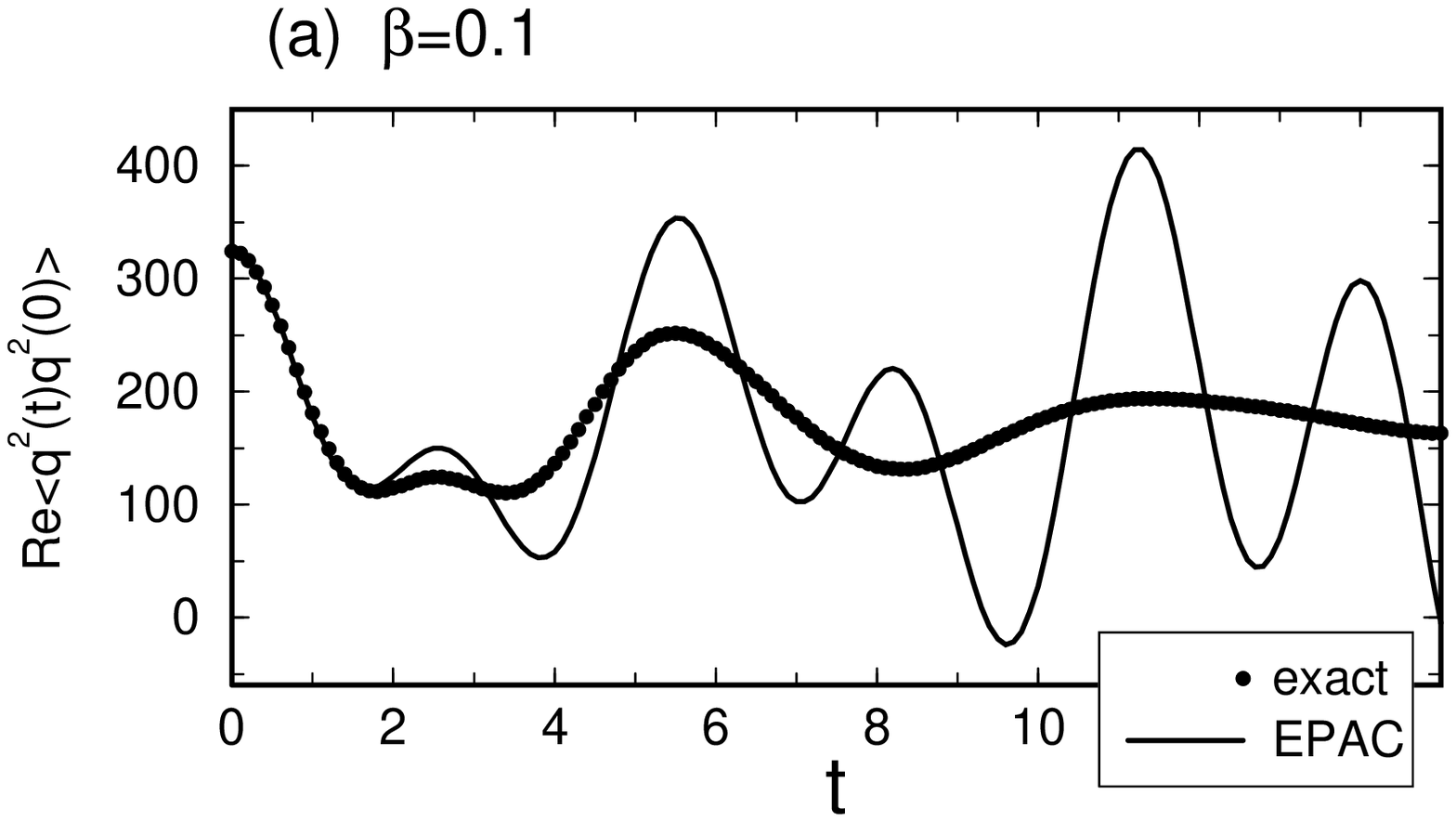}\\
\includegraphics[width=130mm]{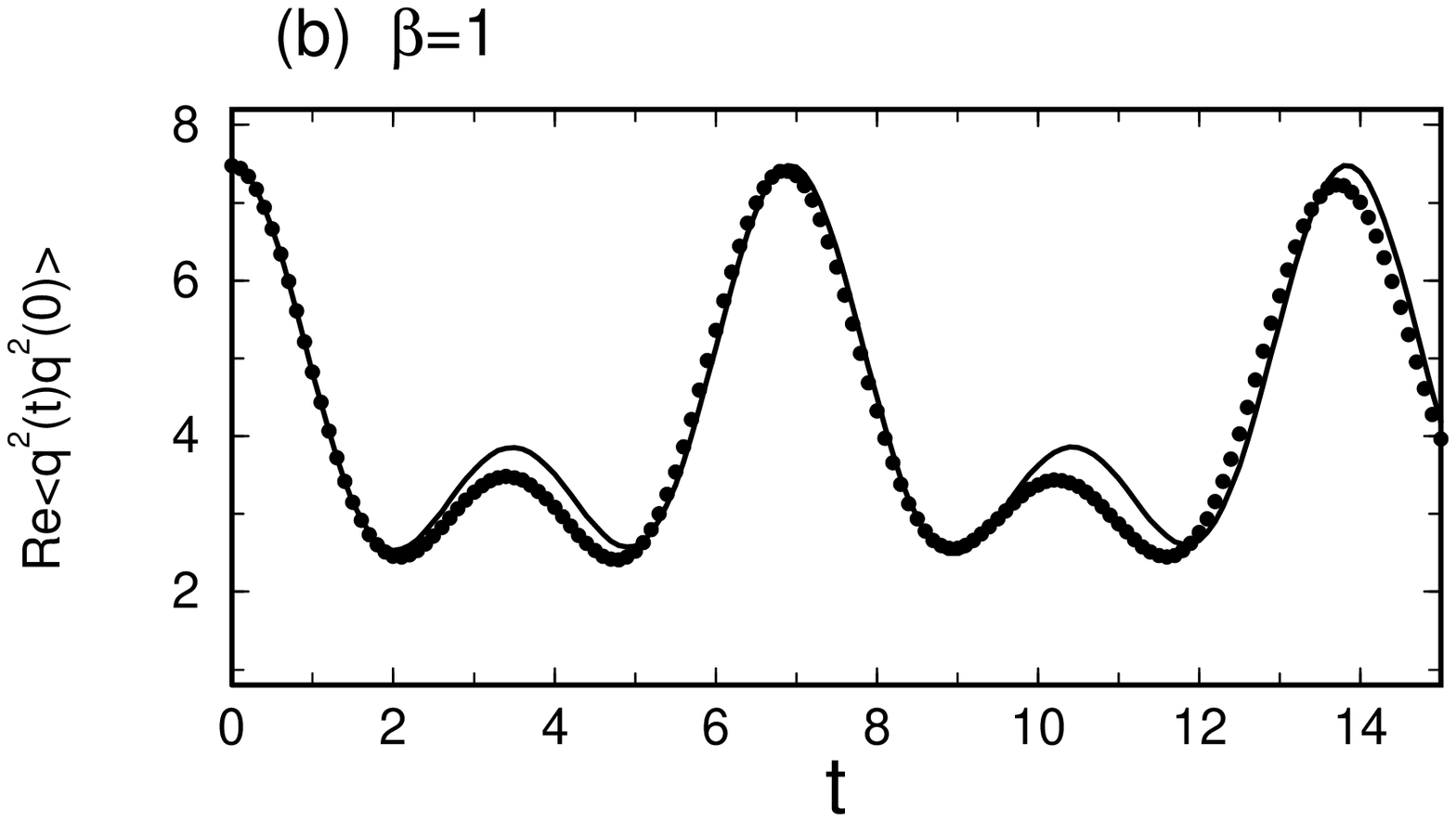}\\
\includegraphics[width=130mm]{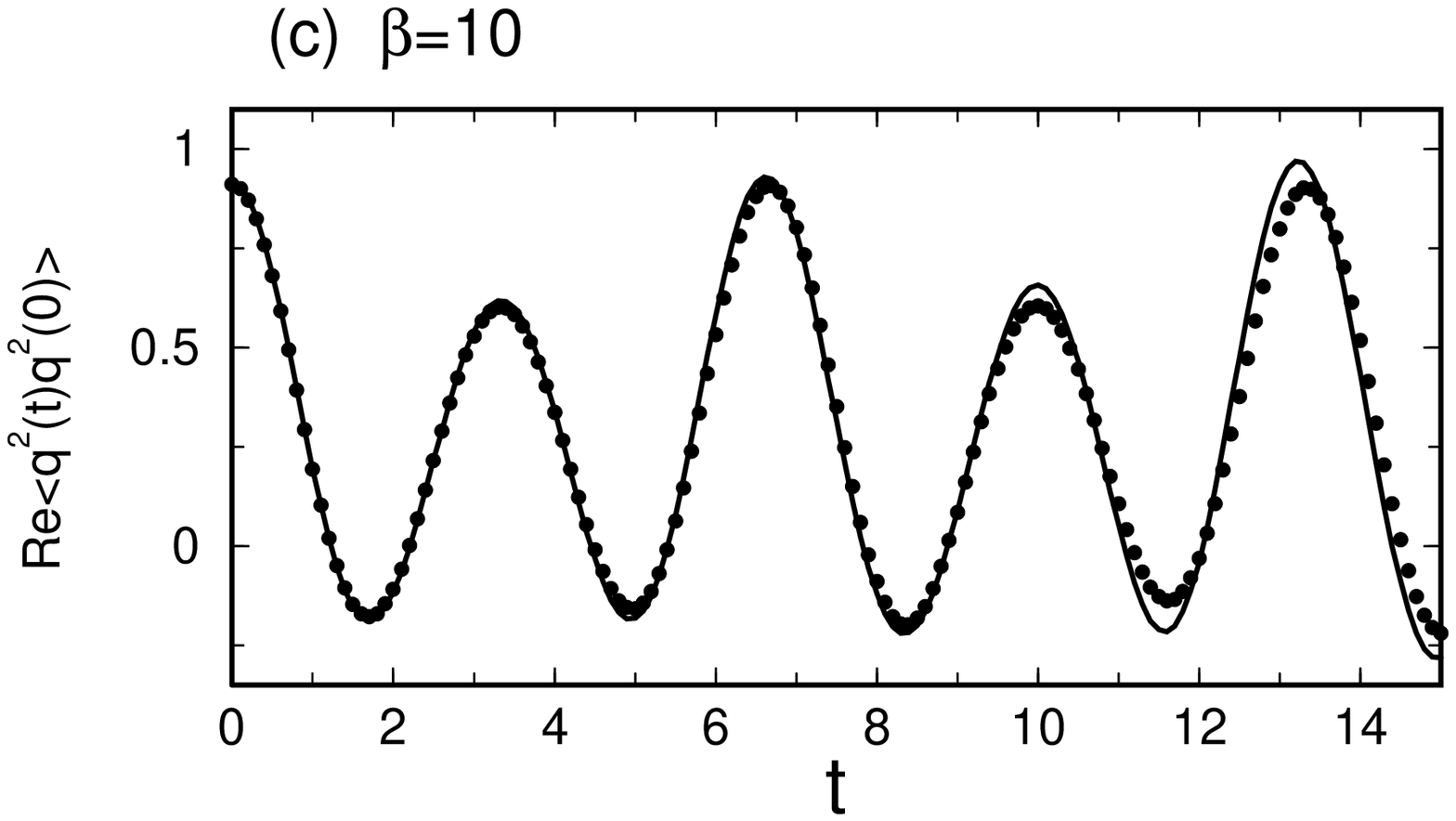}\\
\end{tabular}
\vspace{0mm}
\caption{
The plot of the real part of the exact quantum correlation function 
$\langle \hat{q}^{2}(t)\hat{q}^{2}(0)\rangle_{\beta}$
and  the real part of the EPAC correlation function
$\langle \hat{q}^{2}(t)\hat{q}^{2}(0)\rangle_{\beta}^{\rm EPAC}$
for the quantum anharmonic oscillator (\ref{32}).
(a) at $\beta=0.1$, (b) at $\beta=1$, and (c) at $\beta=10$.
}
\label{fig:aharm}
\end{figure}

Then we have obtained the EPAC correlation function 
$\langle \hat{q}^{2}(t)\hat{q}^{2}(0)\rangle_{\beta}^{\rm EPAC}$
using the quantities listed in Table \ref{tab:table1}.
For reference, 
we have also calculated the exact quantum correlation function 
$\langle \hat{q}^{2}(t)\hat{q}^{2}(0)\rangle_{\beta}$
by solving the Schr\"odinger equation numerically \cite{sch}. 
Figure \ref{fig:aharm} shows the real part of 
$\langle \hat{q}^{2}(t)\hat{q}^{2}(0)\rangle_{\beta}^{\rm EPAC}$
together with the real part of 
$\langle \hat{q}^{2}(t)\hat{q}^{2}(0)\rangle_{\beta}$
at various inverse temperatures $\beta=$ 0.1, 1, and 10.
At $t=0$, each EPAC correlation function is in very good agreement
with the exact correlation function.
This means that the LPA [Eq. (\ref{16})], which  
is the only approximation employed in the EPAC method, 
is fairly good 
for the calculation of the static property 
$\langle \hat{q}^{2}(0)\hat{q}^{2}(0)\rangle_{\beta}$.
In fact, good results have also been obtained in
the calculation of the EPAC correlation function 
involving the linear operator
$\hat{q}$, $\langle \hat{q}(0)\hat{q}(0)\rangle_{\beta}^{\rm EPAC}$,
for the system (\ref{32}) \cite{hk2}.
\par    
On the other hand, as time $t$ increases, 
each EPAC correlation function
deviates from the exact one 
and this deviation
becomes worse at higher temperature.
There are a couple of reasons for such deviation.
The first is the number of oscillation modes
in the correlation functions.
Although the EPAC correlation function (\ref{25})
consists of a limited number of oscillation modes with the frequencies 
$\omega_{\beta}$, $2\omega_{\beta}$, and $3\omega_{\beta}$,
the exact correlation function 
$\langle \hat{q}^{2}(t)\hat{q}^{2}(0)\rangle_{\beta}$,
in general,   
consists of many oscillation modes.
Then at higher-temperature $\beta=0.1$ [Fig. \ref{fig:aharm} (a)],  
a larger number of oscillation modes 
get to contribute to the exact correlation function,
resulting in rapid damping.
The EPAC correlation function 
$\langle \hat{q}^{2}(t)\hat{q}^{2}(0)\rangle_{\beta}^{\rm EPAC}$
including only
three oscillation modes 
cannot represent such damping behavior.
On the other hand, 
as the temperature lowers [Fig. \ref{fig:aharm} (b) and (c)], 
the EPAC correlation function becomes very closer to the exact one
because oscillation modes involved in the exact correlation function  
become fewer.
This is the same behavior as we observed in
the calculation of
$\langle \hat{q}(t)\hat{q}(0)\rangle_{\beta}^{\rm EPAC}$
for the system (\ref{32}) \cite{hk2}.
The second reason for the disagreement between
the EPAC and the exact results at long time
is found in the anomalous terms 
in the EPAC correlation function (\ref{25}).
In Eq. (\ref{25}),
the components $A(t)$ and $B(t)$ both
contain the terms proportional to 
$t\sin 2\omega_{\beta}t$ or $t\cos 2\omega_{\beta}t$,
which exhibit amplified oscillations 
and diverge in the long time limit.
It is expected that 
these anomalous terms disappear
only if we calculate  
the higher order derivative terms 
in the effective action $\Gamma_{\beta}[Q]$
beyond the LPA [Eq. (\ref{16})].
A divergence-free EPAC correlation function 
could also be obtained if  
we omitted all the terms proportional to  
$\tau e^{-2\omega_{\beta}\tau}$ or $\tau e^{2\omega_{\beta}\tau}$
in Eq. (\ref{24}).
However, it should be noted that such prescription
would break the periodicity of 
the imaginary time Green function,
$\langle T \hat{q}^{2}(\beta\hbar)\hat{q}^{2}(0)\rangle_{\beta}
=\langle T \hat{q}^{2}(0)\hat{q}^{2}(0)\rangle_{\beta}$.
\par
Finally we mention how the EPAC method could capture the anharmonic effects
in quantum statistical systems.
In the EPAC method, all the quantum/thermal effects are included via  
the standard effective potential $V_{\beta}(Q)$. 
For the harmonic case (\ref{26}), the effective frequency equals the classical frequency
($\omega_{\rm eff}=\omega$), and the higher-order coefficients vanish
($a_{n}=0$ for $n\geq 3$). However, for anharmonic systems,
the effective frequency $\omega_{\rm eff}$ deviates from the classical one
and the higher-order coefficients $a_{n\geq 3}$ have nonzero values.
As seen in Eq. (\ref{25}), 
the coefficients $a_{n\geq 3}$ contribute 
only to the amplitude of the oscillations of 
the EPAC correlation functions, 
while the effective frequency $\omega_{\rm eff}$
provides oscillation modes with the frequency 
$l\omega_{\beta}$ ($l$: integer).
The EPAC is a method 
to approximate the exact dynamics of quantum systems
using a finite number of oscillation modes with the frequency 
$l\omega_{\beta}$, and therefore 
the EPAC correlation function always exhibits harmoniclike oscillations.
That is, although the anharmonic effects are included via the quantities
$\omega_{\rm eff}$ and $a_{n\geq 3}$,
the EPAC correlation functions cannot reproduce a certain type of anharmonic effects 
such as the dephasing and the rapid damping 
at higher temperature [Fig. \ref{fig:aharm} (a)].
This quasiharmonic property of the EPAC correlation function 
comes from the LPA [Eq. (\ref{16})],
the only approximation employed in the EPAC method.
At the same time, we should note an important problem arising in 
the standard effective potential approach when it is applied to 
quantum systems that include  
the dissociation limit $\lim_{|\bm{q}|\to\infty}V(\bm{q})=0$ \cite{hk2,ct};
for example, the systems represented in terms of  
the  Morse potential $V(r)~(r=|\bm{q}_{1}-\bm{q}_{2}|)$ are 
classified into this category.
For such systems with the dissociation limit,
the minimum should disappear in the standard effective potential
$V_{\beta}(R)$ because it is always convex for $0< R < \infty$
according to its definition \cite{riv,ps,hk1}.
Then the frequency $\omega_\beta$ cannot be defined;
this would lead to some unphysical flaw.
Therefore, the EPAC is not very suitable for the full description of,
e.g., the dissociation reaction of isolated diatomic molecules.
Still we note that at low temperatures 
the EPAC does work well because
the polynomial expansion around the potential minimum $r_{\rm min}$
is so valid as
to neglect the dissociation limit approximately \cite{hk2}.
\par
As for quantum liquids whose intermolecular interaction is 
represented in terms of, e.g., the Lennard-Jones potential with
the dissociation limit as well, the situation of such many-body
systems is not so simple
as an isolated diatomic system,
because potential-energy surface in configurational space
includes many local minima \cite{sw,zw}.
Reichman and Voth have discussed 
various effective harmonic theories 
for liquid dynamics associated with the curvature of 
such potential-energy surface \cite{rv}.
Certainly, there is a similarity between the
effective harmonic theories for liquids and 
the standard effective potential approach we are discussing, 
in that  $\omega_\beta$ is
defined as the second derivative of the standard effective potential 
$V_{\beta}(Q)$ as well.  
However, the standard effective potential and the derived EPAC
method is not suitable for describing the diffusion in liquids 
because some well-defined multidimensional $V_\beta(\bm{Q})$ 
should be convex to have  
a single minimum for one-dimension, resulting in just the oscillatory
motions with real positive frequencies $\omega_\beta^{(i)}$.
Thus, at the present stage, the EPAC method does not fully capture the
molecular diffusion in quantum liquids \cite{bLPA}.  
\par
Rather, the EPAC method should be useful for the investigation of, e.g.,  the 
coherent dynamics in bound systems 
involving  proton transfer reaction \cite{proton}
for which the model potentials are typically represented as 
double-well type such as $V(q)=-aq^{2}+bq^{4} ~(a,b>0)$.
These potentials are asymptotically superlinear,
$\lim_{|\bm{q}|\to\infty}V(\bm{q})/|\bm{q}|> 0$,
without dissociation limit.
In such potential systems the quasioscillating behavior of 
a correlation function is essential. It is true that this oscillation
is a consequence of the quantum coherence between the localized 
states in the local minima of $V(\bm{q})$. The EPAC method based
on $V_\beta(\bm{Q})$ is suitable for describing such oscillation \cite{hk1},
because its frequencies $\omega_\beta^{(i)}$ are evaluated from 
the convex effective potential $V_\beta(\bm{Q})$, which includes 
the effects of quantum interference between the localized states \cite{riv,owy}.

\subsection{Truncated EPAC method}
\hspace*{\parindent}
In this subsection we present a further approximation scheme
useful for practical computation.
Here we treat 
the truncated expansion of $V_{\beta}$
around the minimum $Q=Q_{\rm min}$,
\begin{eqnarray}
V_{\beta}(Q)=\sum_{n=0}^{2}
\frac{a_{n}}{n!}(Q-Q_{\rm min})^{n}
.\label{33}
\end{eqnarray}
This means that 
we approximate the standard effective potential $V_{\beta}$
as an effective harmonic potential 
and omit all 
the higher order derivatives of $V_{\beta}(Q)$,
$a_{n}=\partial^{n}V_{\beta}/\partial Q^{n}|_{Q=Q_{\rm min}}
~(n\geq 3)$.
Since the determination of $a_{n\geq 3}$
requires a heavy computation of $V_{\beta}(Q)$
with high precision, 
the truncation such as Eq. (\ref{33}) reduces 
the computational cost significantly.
\par
\begin{figure}
\begin{tabular}{c}
\includegraphics[width=130mm]{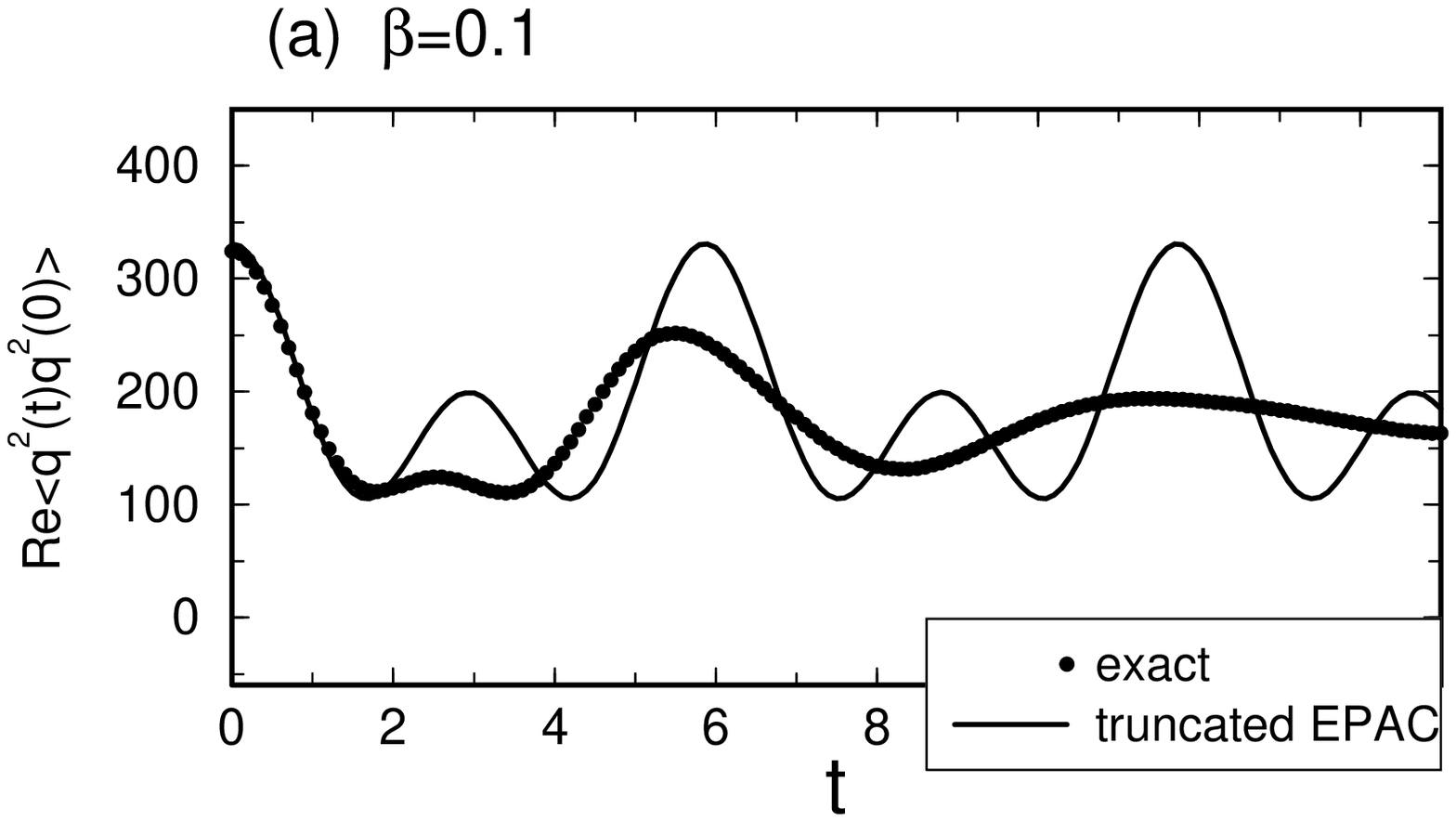}\\
\includegraphics[width=130mm]{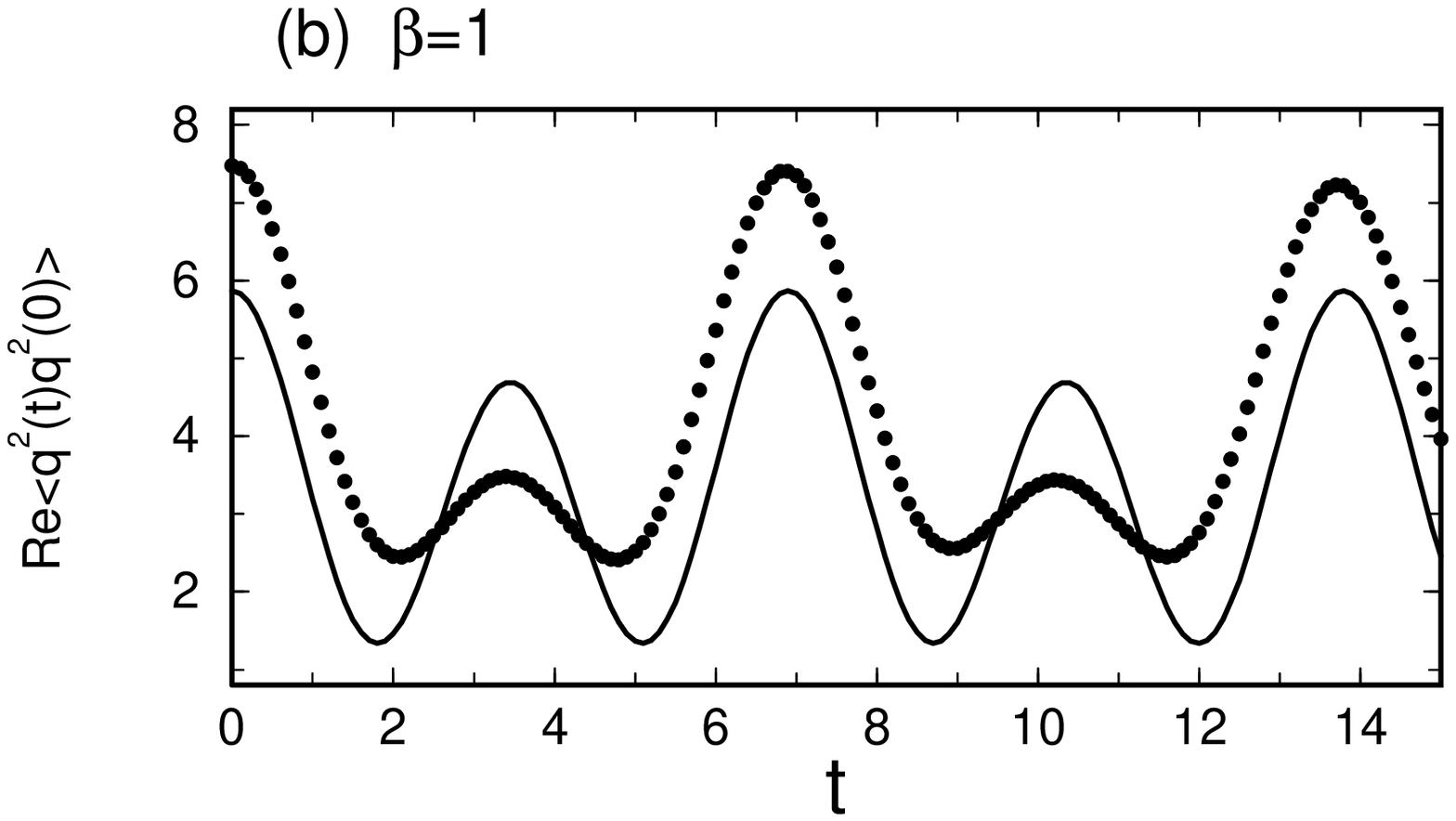}\\
\includegraphics[width=130mm]{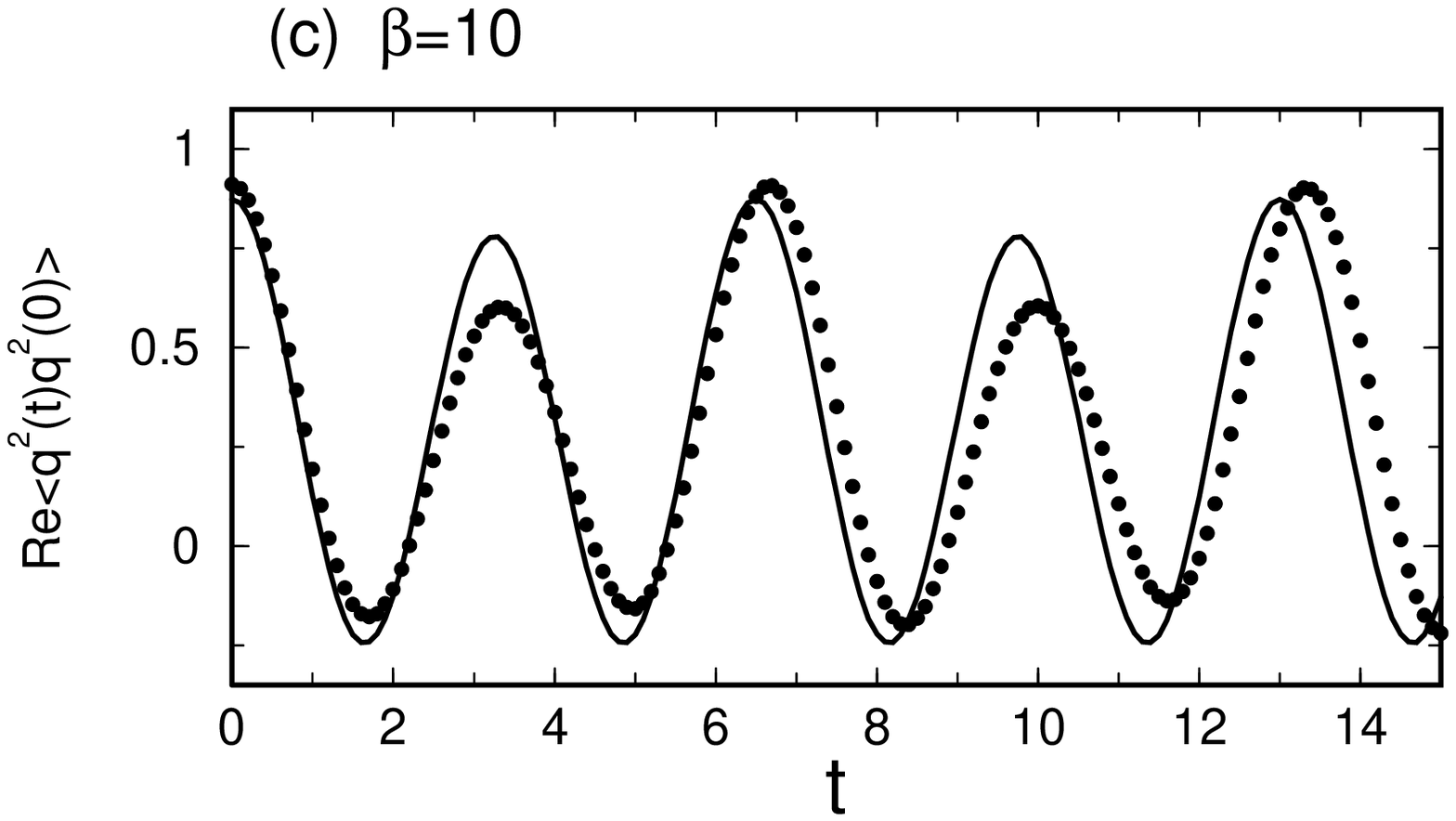}\\
\end{tabular}
\vspace{0mm}
\caption{
The plot of the real part of the exact quantum correlation function 
$\langle \hat{q}^{2}(t)\hat{q}^{2}(0)\rangle_{\beta}$
and  the real part of the truncated EPAC correlation function
$\langle \hat{q}^{2}(t)\hat{q}^{2}(0)\rangle_{\beta}^{\rm tEPAC}$
for the quantum anharmonic oscillator (\ref{32}).
(a) at $\beta=0.1$, (b) at $\beta=1$, and (c) at $\beta=10$. 
}
\label{fig:truncation}
\end{figure}

Using Eq. (\ref{33}), we obtain 
the truncated EPAC correlation function
\begin{eqnarray}
\langle \hat{q}^{2}(t)\hat{q}^{2}(0)\rangle_{\beta}^{tEPAC}
=D(t),\label{34}
\end{eqnarray}
where $D(t)$ is just the function given in Eq. (\ref{25d}).
This is the completely harmonic version of the EPAC method.
Figure \ref{fig:truncation} shows the real part of Eq. (\ref{34})
together with the real part of the exact quantum correlation function
for the quantum anharmonic oscillator (\ref{32}).
The temperatures are the same as in Fig. \ref{fig:aharm}.
Here we have used again the results of $Q_{\rm min}$ and $\omega_{\beta}$
listed in Table \ref{tab:table1}.
\par
We see that, 
at $t=0$, the truncated EPAC correlation functions 
are in good agreement with the exact results 
except for the middle temperature $\beta=1$ [Fig. \ref{fig:truncation} (b)]. 
This behavior can be explained qualitatively as follows.
Since $\coth\alpha\geq 1$,
the highest power of $\coth\alpha$ is regarded 
as the dominating factor
in each term in the full EPAC correlation function (\ref{25}) at $t=0$,
\begin{eqnarray}
a_{4}A(0)&\sim& -\frac{1}{32}\frac{a_{4}}{\omega_{\beta}^{5}}
\coth^{3}\alpha,\label{35}\\
a_{3}^{2}B(0)&\sim&
\frac{5}{96}\frac{a_{3}^{2}}{\omega_{\beta}^{7}}
\coth^{5}\alpha,\label{36}\\
a_{3}Q_{\rm min}C(0)&\sim&
-\frac{1}{3}\frac{a_{3}Q_{\rm min}}{\omega_{\beta}^{4}}
\coth^{2}\alpha,\label{37}\\
D(0)&\sim&
\frac{3}{4}\frac{1}{\omega_{\beta}^{2}}\coth^{2}\alpha,\label{38}
\end{eqnarray}
where $\omega_{\beta}\sim {\it O}(1)$ 
(see Table \ref{tab:table1}).
For lower temperature $\beta=10$,
since $\coth\alpha\simeq 1$ and  
$a_{3}, a_{4}<1$ (see Table \ref{tab:table1}),
all the anharmonic terms in Eq. (\ref{25}) at $t=0$,
i.e., 
$a_{4}A(0)$, $a_{3}^{2}B(0)$, and $a_{3}Q_{\rm min}C(0)$,
become negligible.
Therefore the truncated EPAC result at $t=0$,
$\langle \hat{q}^{2}(0)\hat{q}^{2}(0)\rangle_{\beta}^{\rm tEPAC}$
[Eq. (\ref{34})], is a good approximation to the full EPAC result.
On the other hand, 
as the temperature increases,
$\coth\alpha$ becomes larger than unity, 
and then 
the anharmonic terms 
$a_{4}A(0)$ and $a_{3}^{2}B(0)$
are 
no longer negligible in 
the full EPAC correlation function (\ref{25}),
because they contain
$\coth^{3}\alpha$ and $\coth^{5}\alpha$, respectively. 
Thus the truncated EPAC correlation function 
without the anharmonic terms fails to 
reproduce the full EPAC result
at the middle temperature $\beta=1$ [Fig. \ref{fig:truncation} (b)].
However, 
for higher-temperature $\beta=0.1$,
the coefficients $a_{3}$ and $a_{4}$
become very small 
and therefore the standard effective potential $V_{\beta}(Q)$
has an effectively harmonic shape. 
Consequently, the truncation such as Eq. (\ref{33}) 
becomes a very good approximation to 
the full potential (\ref{20}),
and therefore the truncated EPAC  
can reproduce the full EPAC result very well again. 

\par
As for the long time behavior, a discussion can be made as follows.
The truncated EPAC correlation function 
$\langle \hat{q}^{2}(t)\hat{q}^{2}(0)\rangle_{\beta}^{\rm tEPAC}$
consists of two oscillation modes with the frequencies 
$\omega_{\beta}$ and $2\omega_{\beta}$,
and it lacks the oscillation mode with frequency $3\omega_{\beta}$
which exists in the full EPAC correlation function 
$\langle \hat{q}^{2}(t)\hat{q}^{2}(0)\rangle_{\beta}^{\rm EPAC}$.
Nevertheless, the truncated EPAC correlation function
correctly reproduces the oscillation appearing in the full EPAC 
correlation function at each temperature 
(Figs. \ref{fig:aharm} and \ref{fig:truncation}).
This is because the 
oscillation mode with frequency $3\omega_{\beta}$
exists only in $B(t)$ and its contribution to Eq. (\ref{25})
is relatively small.
It should also be noted that
the truncated EPAC correlation function,
only $D(t)$,
is free from an amplified oscillation expected in Eq. (\ref{25})
and it never diverges in the $t\to\infty$ limit.
We therefore expect that 
the truncated EPAC method, Eq. (\ref{34}), works well 
in low-temperature systems.
Thus the truncated EPAC method can be useful for 
the practical calculation of the nonlinear dynamical properties 
of low-temperature condensed phase systems   
because it should reduce the computational cost.

\section{CONCLUDING REMARKS}
\hspace*{\parindent}
In this paper, we have focused on 
the nonlinear operator problem in quantum dynamics methods.
At first, we have shown how to apply the EPAC method
to the calculation of the correlation function 
$\langle \hat{q}^{n}(t)\hat{q}^{n}(0)\rangle_{\beta}$
and have given the EPAC correlation function 
$\langle \hat{q}^{2}(t)\hat{q}^{2}(0)\rangle_{\beta}^{\rm EPAC}$
as an example.
It has been shown that 
the EPAC method is exact in a harmonic system,
while the other quantum dynamics methods, the CMD and the RPMD, 
fail to reproduce the exact results
even in the harmonic system.  
\par
Then we have applied the EPAC method to the 
asymmetric anharmonic system. 
We have seen that 
the EPAC correlation function at $t=0$ 
agrees well with
the exact correlation function
$\langle \hat{q}^{2}(0)\hat{q}^{2}(0)\rangle_{\beta}$.
As for the long time behavior,
the EPAC result becomes better at lower temperature.
These good properties suggest that 
the EPAC method can be a useful quantum dynamics method 
for nonlinear correlation functions of 
low temperature systems. 
We have also seen that 
the EPAC correlation function contains the terms
which cause amplified oscillation.
We suppose that 
these anomalous terms would disappear 
only if we improved the approximation beyond the LPA.
\par
Finally we have discussed the truncated EPAC method 
which is given by the truncation of 
the standard effective potential $V_{\beta}(Q)$.
We have tested this method in the 
same anharmonic system to which
the full EPAC method has also been applied. 
Although the truncated EPAC method 
is not so good as the full EPAC method,
it works fine at lower temperature. 
Since the truncated EPAC needs only the information up to 
the second order derivative of 
the standard effective potential $V_{\beta}(Q)$,
it should significantly reduce the computational cost.
Therefore, from a practical aspect,
 we expect that the truncated EPAC
is more suitable than the full EPAC 
for complex many-body systems at lower temperature.  
\par
In the present work, the discussions have been restricted in
the evaluation of the correlation function 
$\langle \hat{q}^{n}(t)\hat{q}^{n}(0)\rangle_{\beta}$.
However, the multitime quantum correlation functions, e.g.,
$\langle \hat{q}(t_{1})\hat{q}(t_{2})\hat{q}(t_{3})\hat{q}(0)\rangle_{\beta}$,
are also important property 
especially in the context of the 
nonlinear optical spectroscopy \cite{mukamel}.
It is a challenging task to apply the EPAC method to  
such calculation of multitime quantum correlation functions
in near future.
\begin{acknowledgments}
We are grateful to Ken-Ichi Aoki and Tamao Kobayashi
for their help in the RG method.
This work was supported by a fund for Research and Development 
for Applying Advanced Computational Science and Technology,
Japan Science and Technology Agency (ACT-JST).
\end{acknowledgments}


\newpage
\newpage

\newpage

\newpage


\begin{thebibliography}{99}
\bibitem{bLPA} 
The EPAC method could capture the diffusive motion in liquids
if we calculated  
the higher-order derivative terms 
in the effective action $\Gamma_{\beta}[Q]$
beyond the LPA [Eq. (\ref{16})].
\end{thebibliography}
\end{document}